\documentclass{emulateapj}

\newcommand{\msun}{{M}_{\odot}}

\newcommand{\mdot}{\dot{M}}
\newcommand{\msyr}{\msun \ {\rm yr^{-1}}}

\slugcomment{}

\shorttitle{The White Dwarf Nature of Mira B}
\shortauthors{Sokoloski \& Bildsten}

\begin{document}


\title{Evidence for the White Dwarf Nature of Mira B}


\author{J. L. Sokoloski}
\affil{Columbia Astrophysics Laboratory, Columbia University,
    New York, NY 10027, USA}

\and

\author{Lars Bildsten}
\affil{Kavli Institute for Theoretical Physics and Department of
  Physics, Kohn Hall, University of California, Santa Barbara, CA
  93106, USA} 




\begin{abstract} 

The nature of the accreting companion to Mira --- the prototypical
pulsating asymptotic giant branch star --- has been a matter of debate
for more than 25 years.  Here we use a quantitative analysis of the
rapid optical brightness variations from this companion, Mira~B, which
we observed with the Nickel telescope at Lick Observatory, to show
that it is a white dwarf (WD).  The amplitude of aperiodic optical
variations on time scales of minutes to tens of minutes ($\approx
0.2$~mag) is consistent with that of accreting WDs in cataclysmic
variables on these same time scales.  It is significantly greater than
that expected from an accreting main-sequence star. With Mira~B
identified as a WD, its ultraviolet (UV) and optical luminosities,
along with constraints on the WD effective temperature from the UV,
indicate that it accretes at $\sim10^{-10}\, \msyr$.  We do not find
any evidence that the accretion rate is higher than predicted by
Bondi-Hoyle theory.  The accretion rate is high enough, however, to
explain the weak X-ray emission, since the accretion-disk boundary
layer around a low-mass WD accreting at this rate is likely to be
optically thick and therefore to emit primarily in the far or extreme
UV.  Furthermore, the finding that Mira~B is a WD means that it has
experienced, and will continue to experience nova explosions, roughly
every $10^6$ years.  It also highlights the similarity between Mira~AB
and other jet-producing symbiotic binaries such as R~Aquarii,
CH~Cygni, and MWC~560, and therefore raises the possibility that
Mira~B launched the recently discovered bipolar streams from this
system.

\end{abstract}






\section{Introduction} \label{sec:intro}

Variations in the strength of emission from the accretion region
around a compact object can reveal fundamental properties of the
compact object and its accretion structure.  For example, the
frequencies associated with features in the power density spectra
(PDS) of these brightness variations are linked to physical time
scales at different locations in the accretion disk
\citep[e.g.,][]{mauche02,warner04,done07}.  Such PDS features include
quasi-periodic oscillations, ``breaks'' where the slope of the PDS
changes, and broad-band components that can be fitted with Lorentzian
functions \citep{psaltis99}.  Empirically, the locations of these
features depend on the type of object; in X-ray binaries, the highest
frequency at which significant variability exists is greater for
neutron-star than black-hole accretors \citep{sunyaev00}, consistent
with the higher Keplerian frequency at the inner edge of the accretion
disk around a neutron star.  Even an apparently featureless PDS may
hold the imprint of the underlying accreting object, as is evidenced
by the correlation between the normalization of the PDS at a
particular frequency and the mass of the black hole in active galaxies
compared to X-ray binaries \citep{nikolajuk04,gierlinski08}.

An important case where accretion-related variations can reveal the
nature of an accreting object is the wide, interacting binary-star
system Mira~AB ($o$ Ceti).  Mira~AB is a `symbiotic-like' or `weakly
symbiotic' system at a distance of 107~pc \citep{knapp03}.  According
to the preliminary orbit of \cite{prieur02}, the orbital period is
497.9 years, and the inclination is $112.0^\circ$ (i.e., we see the
binary nearly edge-on).  The donor star is the prototype of the
Mira-type class of pulsating AGB stars, with a pulsation period of
332~days \citep{hoffleit97}.  A more compact companion, whose nature
has been controversial, accretes from the wind of the AGB star.
Although most indicators suggest that the accretor is a WD
\citep[e.g.,][]{warner72,reimers85}, several authors have argued that
the low X-ray luminosity implies it is a main-sequence (MS) star
\citep{jura84, kastner04}.  Determining whether Mira~B is a WD or a MS
star is 
imperative because its nature and evolutionary status impact our
understanding of: 1) the efficiency of wind-fed accretion; 2) mass
transfer through disks with sizes of $\sim 10^{13}$~cm; 3) X-ray
emission from symbiotic stars; and 4) the generation of the bipolar
structure in planetary nebulae (PNe).

Because Mira~AB is spatially resolved in the optical, X-ray, and radio
\citep{karovska97,karovska05,matthews06}, it is an excellent place to
investigate the accretion of a stellar wind from a detached companion.
Roche lobe overflow is thermally unstable when the donor star in a
binary is an evolved giant with a mass in excess of $5M/6$
\citep[where $M$ is the mass of the accretor;][]{webbink83}.  Hence
symbiotic binaries, which typically have low-mass accretors
\citep{mikolajewska03}, 
tend to transfer material via gravitational capture of the red-giant
wind.  In Mira AB, X-ray observations revealed a bridge of material
between the two stars \citep{karovska05} that likely arises from the
gravitational effects, as the wind speed
\citep[6~km~s$^{-1}$;][]{knapp85} is of the same order as the stellar
velocities.  Mira may thus be the first clear example of a new mode of
mass transfer --- dubbed `wind Roche lobe overflow' \citep{mohamed07}
or `gravitational focusing' \citep{devalborro09} --- that lies
somewhere between standard Roche-lobe overflow and Bondi-Hoyle
accretion.  This type of mass transfer may occur in those wide
binaries that ultimately explode as type Ia supernovae \citep[SNIa;
  e.g.,][]{patat07,simon09,blondin09}, with the yield from this
channel of SNIa production depending on the efficiency of the mass
transfer from the RG to the WD.  Since the inferred accretion rate
onto Mira~B ($\mdot = L R/G M$, where $L$ is the luminosity due to
accretion, $R$ is the radius of the accretor, and $G$ is the
gravitational constant) differs by almost two orders of magnitude
depending upon whether Mira~B is a WD or a MS star, determining the
nature of Mira~B is critical 
for estimating $\mdot$ and the ability 
of this channel to produce a SNIa.

Mira~AB will also test the hypothesis that the shapes of PNe are the
result of interactions between a star in its final stages of evolution
and a binary companion \citep{demarco09}.  Observations indicate that
pre-PNe commonly have bipolar morphologies and jets
\citep[e.g.,][]{bujarrabal01,sahai07}, and most models for producing
these structures require a binary companion \citep[e.g.,][and
  references therein]{demarco09,huggins09}.  \citet{miszalski09},
however, estimated that only 10--20\% of PNe contain binary central
stars with separations small enough for the binary to have experienced
a phase of common-envelope evolution, and \citet{huggins09} argued
that the roundness of the majority of the halos of pre-PNe indicates
that any binary companions in these systems must either be quite
distant (as in Mira) or have very low masses.  Mira~A is near its
evolutionary end state, it has a distant binary companion, and recent
ultraviolet (UV) images from the $GALEX$ satellite revealed bipolar
``streams" of knots to the north and south \citep{martin07}.
Therefore, resolution of the issue of the nature of the companion
would clear the path for models of Mira~AB to confront questions such
as how wide binaries generate bipolar structure in PNe, which stellar
component launches the bipolar outflows, and when asymmetry first
appears.

In this paper, we use fast $B$-band photometry to show that Mira B is
a WD.  We describe our rapid photometric observations in
\S\ref{sec:obs}, and the resulting light curves and power density
spectra in \S\ref{sec:results}.  In \S\ref{sec:disc}, we explain how
the amplitude of the optical variations on time scales of minutes
reveal Mira~B to be a WD.  We also estimate a characteristic accretion
rate onto Mira~B and reinterpret the X-ray emission.  We summarize our
conclusions and explore several implications in
\S\ref{sec:conclusions}.

\section{Observations and data reduction} \label{sec:obs}


\begin{deluxetable}{cccccc}
\tabletypesize{\footnotesize}
\tablecaption{Log of Mira~AB Observations \label{tab:obslog}}
\tablewidth{0pt}
\tablehead{
\colhead{Date,} & \colhead{Obs. Start} &
\colhead{Duration} & \colhead{$t_e$} & \colhead{$\Delta t$} &
\colhead{Num. } \\
\colhead{U.T.} & \colhead{(MJD)} & \colhead{(hr)} & \colhead{(s)} &
\colhead{(s)} & \colhead{Pts}     
}
\startdata
1997 Sep 2 & 50693.350 & 3.6 & 25.0 & 46.996 & 256 \\
1997 Nov 3 & 50755.410 & 1.6 & 50.0& 70.0 & 81 \\
1998 Aug 19 & 51044.482 & 1.0 & 18.0 &38.990 & 97 \\
1998 Aug 20 & 51045.381 & 3.5 & 30.0 & 49.988 & 248 \\
1998 Sep 17 & 51073.352 & 4.6 & 25.0 & 46.0 & 355 \\
\enddata
\tablecomments{Obs. Start is the start time for a given observation.
  Duration is the length of the observation.  Columns 4 and 5 lists the
  exposure time and the time between exposure starts 
  as $t_e$ and $\Delta t$, respectively.}
\end{deluxetable}

As part of a photometric survey of 35 symbiotic stars
\citep[see][hereafter SBH]{sokoloski01}, we observed Mira AB with the
1-m Nickel telescope at UCO/Lick Observatory on Mt.~Hamilton, near San
Jose, CA, on 5 nights between 1997 September and 1998 September.
Table~\ref{tab:obslog} contains a log of the observations, each of
which consisted of a series of exposures with an unthinned, $2048
\times 2048$ Loral CCD (referred to at Lick Observatory as `CCD 2')
that had 15-$\mu$m pixels and a $6\arcmin\!.3 \times 6\arcmin\!.3$
field of view.  We chose exposure times, $t_e$, of 18 to 50~s to
maximize the signal-to-noise ratio (S/N) while avoiding saturation of
the program star or the one bright comparison star in the field of
view (HD~14411).  Keeping the exposure times at or above 18~s also
kept the observing efficiency ($t_e/\Delta t$, where $\Delta t$ is the
time between observation starts) from dropping much below $\sim 50$\%
and minimized scintillation noise, which can dominate the error at
very short exposure times (SBH).  Including chip pre-processing and
readout times of between 20 and 22 seconds, $\Delta t$ ranged from 39
to 70 seconds.  To obtain evenly spaced data points within a given
observation, and thereby produce data that were compatible with
standard fast Fourier transform (FFT) routines, we employed a timing
system developed especially for this project by W. Deitch (UCO/Lick).
The resulting light curves spanned from 1.1 to 4.6 hours.  To minimize
the contribution of the red giant while still obtaining adequate flux
from the bluer hot component to generate a high S/N on time scales of
minutes, we used a Johnson $B$ filter and observed only near the
minima of the 332-day Mira pulse period, as displayed in Figure~1.

\begin{figure*}
\epsscale{0.9}
\plotone{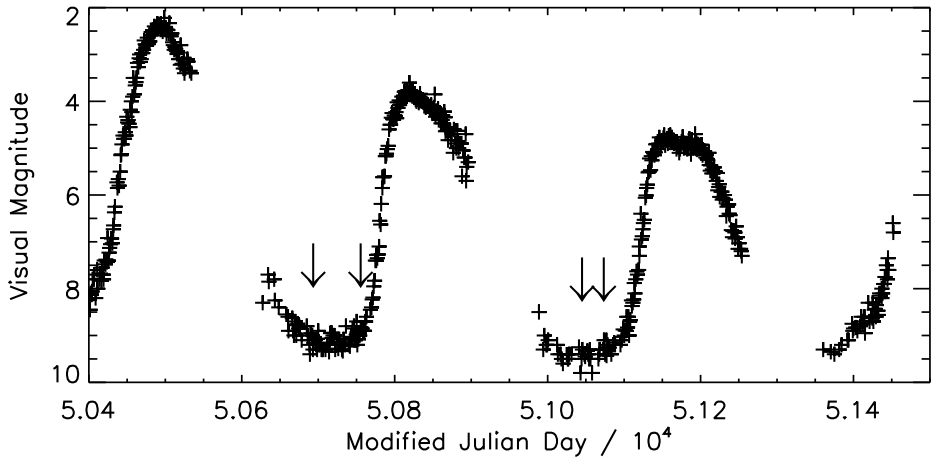}
\caption{Long term visual light curve, from 1996 to 1999, from the
  American Association of Variable Star Observers (AAVSO), with the
  dates of our high-time-resolution $B$-band observations marked with
  arrows. Our observations all occurred near a photometric minimum of
  the pulsating red giant. Since we performed our third and fourth
  observations on subsequent nights (1998 August 19 and 20), they
  appear as a single arrow (the third from the left). \label{fig:miraltlc}}
\end{figure*}

For each observation, we reduced the CCD images using standard
techniques and performed aperture photometry, as described in SBH.
For each image, our reduction included subtracting the electronics
zero point and bias pattern and dividing by an average of multiple
images of the sky, using IDL software based upon IRAF routines
\citep[see][]{gil92,mas97}.  We produced differential light curves
with respect to either the one bright comparison star in the field or
a weighted, ensemble average of 2 -- 4 comparison stars (depending
upon whether the additional comparison stars improved the S/N enough
to warrant the additional lost points due to cosmic rays).  We
estimated the background for each star from an annulus around that
star.  We omitted data points contaminated by radiation events
(`cosmic rays'), points with extremely high background, and data taken
during exceptionally poor weather conditions. None of the comparison
stars showed any evidence for intra-night variability.

\section{Analysis and results} \label{sec:results}

\begin{figure*}
\epsscale{0.9}
\plotone{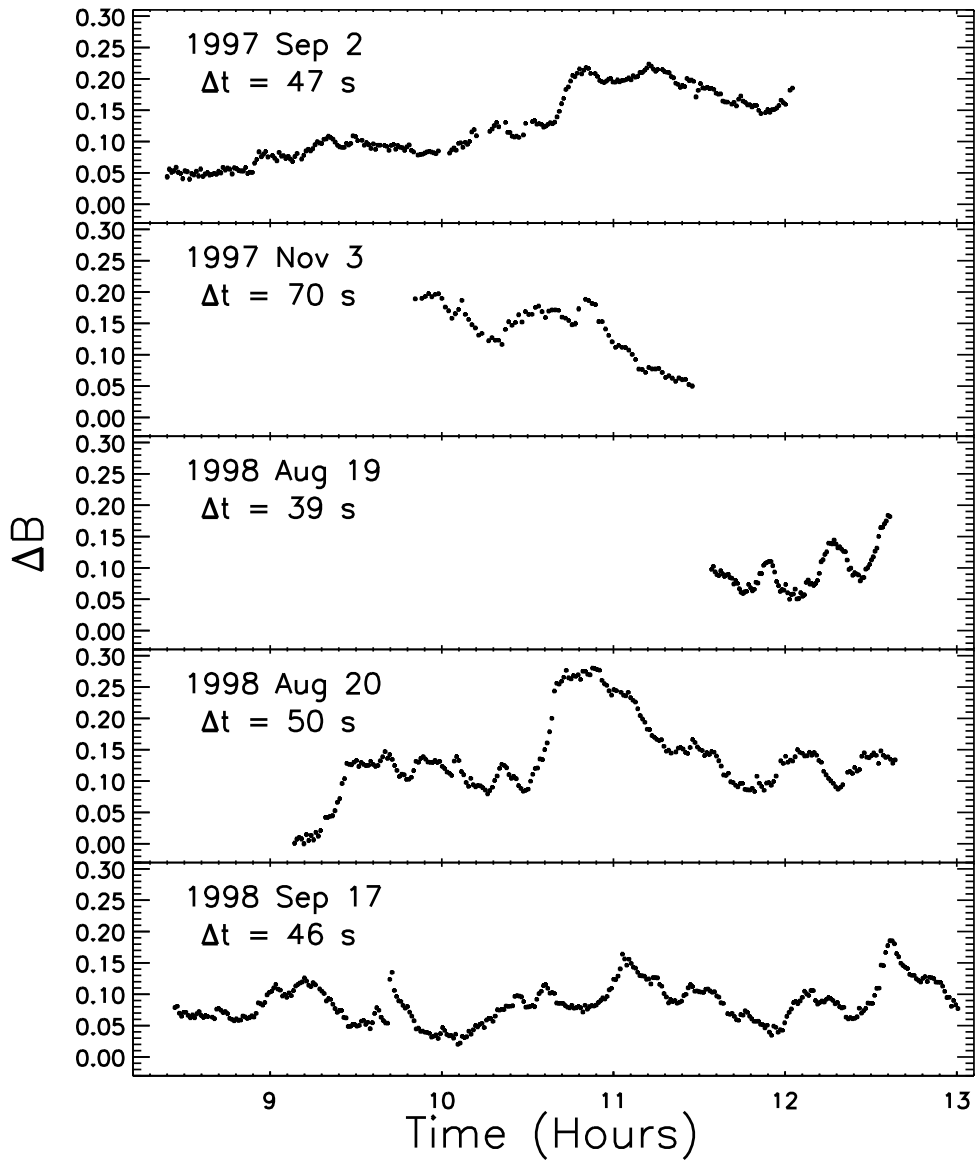}
\caption{Differential $B$-band light curves for Mira~AB.  The times
  between data points for the light curves are $\Delta t = $47, 70,
  39, 50, and 46~s, respectively.  The size of the error bars is
  comparable to that of the point symbols. The time is U.T. on each
  date. \label{fig:miralcs}}
\end{figure*}

\begin{figure}
\epsscale{1.22}
\plotone{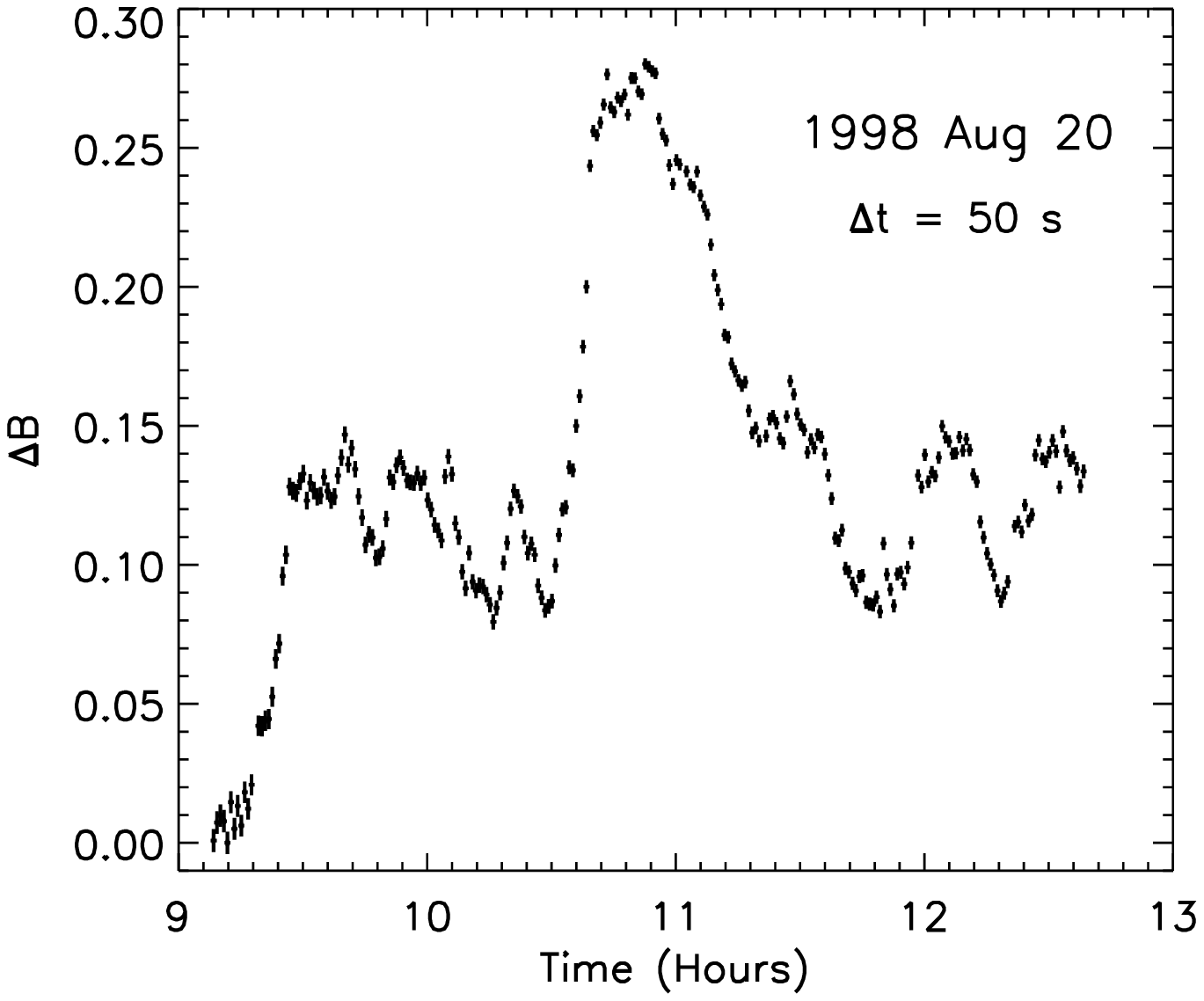}
\caption{Close-up view of the differential $B$-band light curve for
  Mira~AB on 1998 August 20. The size of the error bars is comparable
  to that of the point symbols. The time is U.T. \label{fig:miraexlc}}
\end{figure}

Rapid aperiodic variability, like the variability from cataclysmic
variables that is often termed `flickering' \citep[CVs;
  e.g.,][]{warner95,bruch00}, is evident on all 5 nights.
Figure~\ref{fig:miralcs} shows the differential light curves from the
5 observations.  On each night, the measured root mean square (rms)
variation, $s$, exceeded the rms variation expected from the known
errors, $s_{\rm exp}$, by more than an order of
magnitude. Table~\ref{tab:pdsindices} lists $s$ and $s_{\rm exp}$ (the
estimation of which is discussed in detail in SBH) for each
observation.  Figure~\ref{fig:miralcs} demonstrates that the $B$-band
flux can change by more than 5\% ($\approx 50$~mmag) in less than 5
minutes and more than 25\% ($\approx$0.25~mag) in less than 2 hours.
Taking into account the non-variable light from the red giant, which
contributes $\approx 50$\% of the flux at $B$-band \citep[since Mira~A
  and B had similar brightness in the $HST$ F501N filter when Mira~A
  was near pulse minimum in 1995;][]{karovska97}, these rapid
fluctuations correspond to variations of approximately 10 -- 50\% in
the $B$-band light from Mira~B.  In summary, correcting for the
contribution from the red giant, the $B$-band flux from Mira~B varied
as fast as a few per cent (or tens of mmag) per minute, and by order
unity in tens of minutes, as exhibited in the light curve from 1998
August 20 (Figure~\ref{fig:miraexlc}).

\begin{deluxetable}{cccccc}
\tabletypesize{\footnotesize}
\tablecaption{Optical Variability of Mira~AB\label{tab:pdsindices}}
\tablewidth{0pt}
\tablehead{
\colhead{Date} & \colhead{ $s_{exp}$} & \colhead{$s$\tablenotemark{a}}
& \colhead{Powerlaw} &
\colhead{Freq Range} & \colhead{$\chi^2_{\nu}$} \\
\colhead{(U.T.)} & \colhead{(mmag)} &
\colhead{(mmag)} & \colhead{Index} &
\colhead{(mHz)} & \colhead{}}
\startdata
1997 Sep 2& 2.7 & 50.9 & $-1.8\pm0.2$ & 0.08 -- 3.6 & 0.8 \\*
1997 Nov 3& 2.6 & 37.9 & $-2.4\pm0.3$ &0.2 -- 3.4 & 0.8 \\* 
1998 Aug 19& 2.6 & 29.3 & $-2.1\pm0.2$ & 0.3 -- 9.3 & 0.7 \\* 
1998 Aug 20& 2.3 & 55.3 & $-2.0\pm0.1$ & 0.08 -- 5.5 & 0.9 \\* 
1998 Sep 17& 2.5 & 28.7 & $-1.9\pm0.1$ & 0.06 -- 10.8 & 0.9 \\ 
\enddata
\tablenotetext{a}{The measured rms variation, $s$, includes variations
at all frequencies to which a given observation was sensitive; the
range of frequencies sampled for each observation is approximately
equal to the range over which we fitted the powerlaw model, shown
in column 5.}
\end{deluxetable}

\begin{figure*}
\epsscale{.90}
\plotone{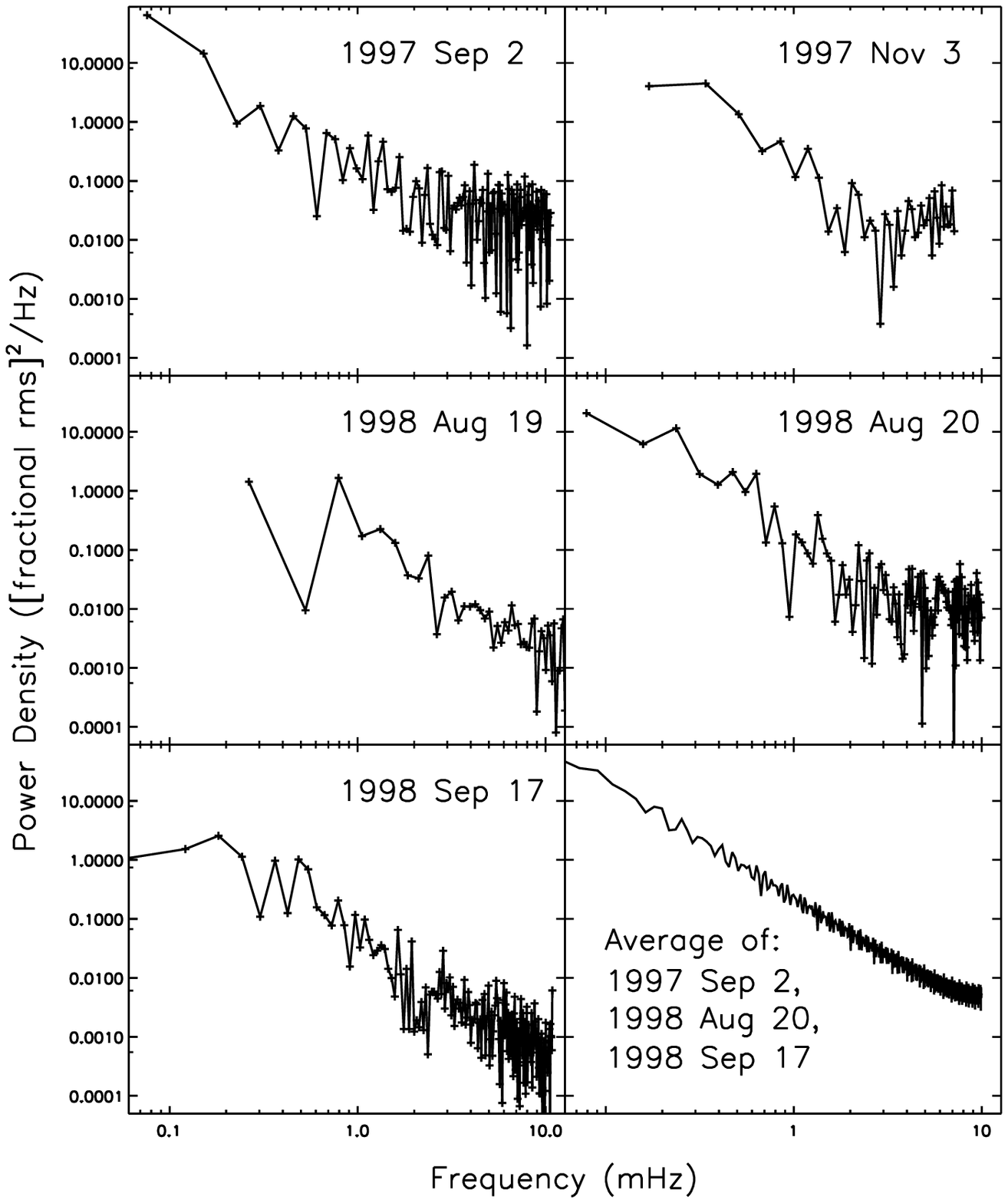}
\caption{Power spectra of the light curves shown in Figure
  \ref{fig:miralcs}.  The average PDS in the lower right panel was
  produced by padding the light curves with enough zeros to generate
  frequency spacings that were the same for each of the nightly
  PDS.\label{fig:mirapdss}}
\end{figure*}

The PDS of all 5 light curves reveal broadband power with a powerlaw
shape --- $P_\nu \propto \nu^\alpha$ (where $P_\nu$ is the power
density and $\nu$ is the frequency) --- between approximately 0.1 and
10 mHz (see Table~\ref{tab:pdsindices}).  Since our data points were
evenly spaced, we used the IDL routine {\em fft} to compute the PDS.
We normalized the PDS to the variance of the light curves using the
Miyamoto normalization \citep{miyamoto91}, in which the square root of
the integrated power between two frequencies is equal to the
fractional rms variation between those two frequencies.  For PDS with
a powerlaw index $\alpha = -2$, the square root of the power at a
given frequency equals the fractional rms variation summed over all
frequencies above that frequency.  Figure~\ref{fig:mirapdss} shows the
PDS corresponding to the light curves in Figure~\ref{fig:miralcs}.
Table~\ref{tab:pdsindices} lists the powerlaw indices for each
observation; they ranged from -1.8 to -2.4.  The weighted average of
the 5 powerlaw indices is $\alpha = -1.969 \pm 0.004$.  Since this
value is very close to $\alpha = -2$, we refer to the stochastic
variations from Mira as {\em red noise}.

Extrapolating the PDS where necessary using a power-law index of
$\alpha = -2$, and taking into account the constant light from the red
giant, we find that the average rms variation from Mira~B between 0.1
and 10 mHz is approximately 9\%.  We derived this value from a
measured rms variation for the total $B$-band light in this same
frequency range of 4.5\%, since emission from the red giant, which
does not vary at these frequencies, constitutes approximately half of
the $B$-band flux.  Thus, the $B$-band rms variation from Mira~B is at
least twice the rms variation of the total $B$-band light.  In the
restricted frequency range of 1 -- 10~mHz, which is useful for
comparison with published rms variations in this frequency range for
CVs, the average rms variation from Mira~B is 3\% (correcting a
measured rms variation of 1.5\% in this frequency range for the
contribution from the red giant).

We did not find any evidence for coherent, periodic variations on time
scales of minutes to hours in the optical brightness of Mira~B.
Taking the broadband power into account \cite[as discussed in the
  Appendix of][]{sokokenyon03}, no peaks in the PDS rose significantly
above the level of the red noise.

\section{Discussion} \label{sec:disc}

%

The nature of the accreting companion to Mira~A may hold the key to
improved understanding of mass transfer via a wind, as well as the way
in which binaries generate asymmetrical structure in PNe.  Whereas
early observations of rapid CV-like optical variability by
\citet{walker57} and \citet{warner72} originally led most authors to
assume that Mira~B was a WD, later revelations that the X-ray
luminosity was lower than expected motivated \citet{jura84} and then
\citet{kastner04} to propose that Mira~B was instead an accreting MS
star.  Though \citet{ireland07} reported some marginally significant
(4-$\sigma$) features in the optical spectrum that appeared to suggest
a MS companion, the UV spectrum is much more indicative of a WD
\citep{reimers85}.  Below, we use a quantitative examination of the
amplitude of the optical variability on time scales of minutes, along
with an estimate of the expected optical depth of the accretion-disk
boundary layer, to show that the optical to X-ray observations are
more consistent with a WD accretor.

\subsection{Rapid optical variations and the nature of Mira~B}

The persistent presence of minute-time-scale stochastic optical
variations with the observed amplitude is a strong indicator that Mira
B is a WD.  Qualitatively, whereas we are not aware of a single
example of an accreting MS star with a high time resolution optical
light curve resembling those of Mira, most accreting WDs have optical
light curves that show Mira-like aperiodic variations with a time
scale of minutes --- i.e., faster than the dynamical time of a MS
star.  Quantitatively, the PDS of our Mira light curves have shapes
and strengths consistent with those of WDs and not MS stars.  The PDS
from Mira has a powerlaw index of approximately $-2$, like the
high-frequency PDS of other accreting objects \citep[including WDs,
  X-ray binaries, and AGN; e.g.,][]{bruch92,done07,gierlinski08}.  We
therefore take the rapid variability to be related to accretion.
Furthermore, the rms amplitude of the $B$-band brightness variations
from Mira B between 1 and 10 mHz (after accounting for the
contribution from the red giant) is around $3$\% --- almost identical
to the characteristic value for CVs, which \citet{barros08} found to
have rms variabilities in the same frequency range of 2.4\% in the
$g^\prime$ and $r^\prime$ bands and 4.5\% in the $u^\prime$ band.

Theoretical considerations suggest that we should not see such rapid
variations from a MS accretor. The dynamical and viscous times at the
inner edge of an accretion disk around a MS star are $\approx 100$
times longer than those at the inner edge of a disk around a WD. Since
these are likely the relevant time scales for setting the speed of
brightness fluctuations, we expect an accreting MS star to exhibit
stochastic variations that are stronger at low frequencies and weaker
at high frequencies than the brightness fluctuations from an accreting
WD.  In fact, light variations have been detected from the accreting
MS stars in Algol binaries, and these variations preferentially occur
on time scales of days rather than minutes; \citet{olson95} found
variations in the strength of double-peaked H$\alpha$ emission, which
they argue trace changes in the inner accretion disk, on time scales
of days in a sample of 9 Algol-type binaries.  Although we have no
empirical constraints on variations from a MS-star disk in the
frequency range of 1 -- 10~mHz, we expect the amplitude of
fluctuations in this frequency range to be orders of magnitude smaller
than those from accreting WDs, if, for example, the PDS from an
accreting MS star is simply that of a WD but shifted to lower
frequencies.  Therefore, given that the amplitude of the
high-frequency fluctuations from Mira are consistent with those from
CVs, and inconsistent with expectations for a MS star, we interpret
the accreting object in Mira as a WD.

\subsection{Reinterpretation of the X-ray emission} 

Before addressing the X-ray emission, we use the luminosity of Mira~B
in the UV and optical to estimate a characteristic accretion rate,
$\mdot$, onto Mira~B.  UV observations with the $IUE$ satellite on
1980 June 16 (when Mira~A was at pulse minimum and therefore did not
contribute to the UV flux) revealed a UV continuum luminosity between
1800 and 3200\AA\, of $L_{UV} \approx 4 \times 10^{32}\; (d/107\,{\rm
  pc})^{2}$~erg~s$^{-1}$ \citep[which we estimated from the top panel
  of Fig.~1 of][]{reimers85}.  As noted by \citet{reimers85}, this
estimate is a lower limit to the UV luminosity, since an accretion
disk could also radiate in the Lyman continuum, which $IUE$ would not
detect.  The $HST$ spectra of Mira~A and Mira~B in Figure 2 of
\citet{karovska97}, taken on 1995 December 11, shows that Mira~B
produced $L_{opt} \approx 4 \times 10^{32}\; (d/107\,{\rm
  pc})^{2}$~erg~s$^{-1}$ between roughly 3200 and 5000\AA. Given that
Mira~B also emits at wavelengths longer than 5000\AA, we estimate that
the total, accretion-powered luminosity of Mira~B is typically at
least $L \approx 10^{33}\; (d/107\,{\rm pc})^{2}$~erg~s$^{-1}$.  This
luminosity implies that Mira~B accretes at a rate of at least
$\mdot_{\rm WD} \approx 10^{-10}\,\msyr \left(L/10^{33}\,{\rm
  erg\,s}^{-1}\right) \left(R/10^9\,{\rm cm} \right) \left(
M/0.6\,\msun\right)^{-1}$.

Consistent with our contention that Mira~B is a WD, an accreting WD
can naturally produce a UV spectrum like the one seen from Mira.  The
$IUE$ spectra from Mira between 1979 and 1983 contained lines with a
range of ionization states, including \ion{N}{5}~1240\AA ,
\ion{C}{4}~1550\AA, and \ion{Si}{4}~1400\AA, and line widths --- from
135~km~s$^{-1}$ for \ion{Fe}{2} absorption, to few hundred km~s$^{-1}$
for \ion{C}{4}~1550\AA\, to $\sim 1,000$~km~s$^{-1}$ for Ly$\alpha$
and \ion{Mg}{2} \citep{reimers85}.  Moreover, the Ly$\alpha$ profiles
that \citet{wood02} and \citet{wood04} reconstructed from H$_2$
fluorescence lines observed by $HST/STIS$ on 1999 August 2 and by
$FUSE$ on 2001 November 27 had full widths at half maximum (FWHM) of
approximately 1,100 to 1,200~km~s$^{-1}$.  Velocities of greater than
$1,000$~km~s$^{-1}$ are larger than the fastest Kepler velocities in
the vicinity of an accreting MS star, but comparable to the Kepler
velocities near an accreting WD.  In fact, line widths of $\sim
1,000$~km~s$^{-1}$ are often seen in the UV spectra of CVs, which
generally emanate from accretion-disk winds
\citep[e.g.,][]{mauche91,froning03}.  Furthermore, \citet{reimers85}
determined that the temperature of the material producing the
high-ionization resonance lines in the $IUE$ spectrum of Mira (which
they take to be the coronal region of the disk) was on the order of
$10^5$~K --- very reasonable for material near an accreting WD.

Regarding the X-ray emission from Mira, an X-ray luminosity that is
several orders of magnitude below the optical luminosity from the
accretion disk, as we see from Mira~B, is exactly what one would
expect if the boundary layer (BL) around the accreting WD is optically
thick \citep{patterson85} or when the spreading layer (SL) on the
accreting WD does not extend above the thin disk \citep{piro04}.  For
a slowly rotating WD, roughly half of the energy released via
accretion is radiated by the BL \citep{shakura73,lyndenbell74} or in
the latitudinal spreading of the material after it arrives on the WD
\citep{piro04}.  If the material is shocked as it arrives, the the
high Keplerian velocities near the surface of the WD produce hard
X-rays that escape if the plasma is optically thin.  With an accretion
rate of $\mdot_{\rm WD} \sim 10^{-10}\, \msyr$, we would expect a
luminosity of at least a few times $10^{32}$~erg~s$^{-1}$ from the BL
around Mira~B.  Estimates of the X-ray luminosity, however, all
indicate that $L_x$ is no more than a few times $10^{29}$~erg~s$^{-1}$
\citep{jura84,kastner04,karovska05}.  Although a detailed
investigation of the X-ray emission is beyond the scope of this paper,
the low $L_x$ most likely reveals that the boundary (or spreading)
layer is optically thick and therefore emitting primarily in the far-
or extreme-UV instead of in the X-rays.  \citet{wheatley03} found that
for the WD in SS~Cygni, the BL became optically thick when the
accretion rate rose above $10^{-10}\, \msyr$, in accord with our
hypothesis that the accretion rate onto Mira~B is high enough to
quench the X-ray emission.  Since the accretion rate at which the BL
transitions from optically thin to optically thick increases with WD
mass \citep{popham95}, the optical thickness of the BL around Mira~B
suggests that the mass of Mira~B is lower than that of the WDs in,
e.g., the symbiotic binaries T~CrB, RT~Cru, and SS73~17, which contain
WDs with masses greater than 1.0~$\msun$ and BLs that are optically
thin at accretion rates as high as a few times $10^{-9}\, \msyr$
\citep{luna07,kennea09,eze10}.

The existence of other WDs that accrete from the wind of a red giant
and produce a very low X-ray luminosity supports our claim that the
X-ray luminosity from Mira~B is consistent with WD accretion.  For
example, the accretor in R~Aqr, which \citet{jura84} also suggested
was a MS star based on its low X-ray luminosity, was later shown by
$IUE$ to have an effective temperature of 61,000~K and the radius of a
WD \citep{meier95}.  More generally, \citet{murset97} found that the
bolometric luminosities of the wind-fed WDs in symbiotic stars (such
as EG~Andromedae, PU~Vulpecula, and RX~Puppis) often exceed their
X-ray luminosities by as much as 5 orders of magnitude.

Finally, a different suggestion for Mira~B --- that the faint X-rays
are coronal emission from a magnetically active, rapidly rotating
low-mass MS star \citep{kastner04} --- cannot explain either the
strong UV emission (compared to X-rays) or the red-noise-type optical
variations.  Whereas the luminosity of the rapidly variable component
of the optical emission from Mira~B is on the order of
$10^{32}$~erg~s$^{-1}$, the time-averaged power of optical radiation
from stellar flares rarely exceeds $10^{30}$~erg~s$^{-1}$
\citep{shakhovskaya89}.  More importantly, a MS star producing an
X-ray luminosity of a few times $10^{29}$~erg~s$^{-1}$ due to coronal
emission would only have a time-averaged optical power from flares of
$\sim 10^{29}$~erg~s$^{-1}$ \citep[][]{shakhovskaya89}.  We can thus
rule out coronal emission from a magnetically active, rapidly
rotating, MS star on energetic grounds.  The radio emission from
Mira~B is also one to three orders of magnitude greater than expected
from a magnetically active dwarf \citep{matthews06}.  In contrast, all
available observations are consistent with the picture in which a WD
is accreting from the wind of the red giant.

\section{Conclusions and implications} \label{sec:conclusions}

Our main conclusion is that the rapid optical variability from Mira~AB
reveals Mira~B to be a WD.  The amplitude of the optical brightness
fluctuations from Mira~B on time scales of minutes is the same as
those from accreting WDs in CVs, and significantly larger than one
would expect from an accreting MS star.  The UV spectra --- especially
the ionization states and widths of the lines --- are consistent with
the conclusion that Mira~B is a WD.  The X-ray luminosity and spectrum
are also consistent with this conclusion if the accretion-disk
boundary layer is optically thick, as one would reasonably expect for
our estimated accretion rate of $\sim 10^{-10}\, \msyr$ \citep[e.g.,
  see][]{wheatley03}.

This accretion rate, which we infer from the optical and UV
luminosities of Mira~B, is consistent with constraints on the
effective temperature of the WD from UV spectra.  Sion (1999) noted
that prolonged accretion will re-heat an otherwise cooling WD, making
it potentially visible in the UV or optical continuum.  Such
rejuvenated WDs are seen in dwarf novae systems during quiescence,
revealing effective temperature of $T_{\rm eff}=$ 10,000--20,000~K
\citep{townsley09}.  This WD re-heating \citep{townsley03,townsley09}
and the resulting $T_{\rm eff}$ depend on both $M$ and the
time-averaged accretion rate, $\langle \dot M\rangle$, thereby
providing another constraint for our analysis of Mira B. From the UV
continuum flux upper limit $F_\lambda (1250 \AA) < 10^{-14} \, {\rm
  ergs \ cm^{-2} \ s^{-1}}$, Reimers and Cassatella (1985) found
$T_{\rm eff}<14,000\, {\rm K}$ for $R=0.012R_\odot$ (for a WD with
$M=0.6M_\odot$ and a distance of 77 pc; a slightly higher $T_{\rm
  eff}$ is allowed at our adopted distance of 107 pc).  A direct
application of \citet[][see their Figure 1]{townsley03} converts this
upper limit on $T_{\rm eff}$ into the long-term accretion rate
constraint $\langle \dot M\rangle< 2 \times 10^{-10}M_\odot \ {\rm
  yr^{-1}}$, in agreement with our earlier estimate.

At this accretion rate ($\dot M\approx 10^{-10}M_\odot \ {\rm
  yr^{-1}}$), the accumulated matter on the WD will burn unstably and
cause a classical novae, leading to the violent ejection of $\sim
10^{-4}M_\odot$ of material at $>1000 \ {\rm km \ s^{-1}}$.  This
ejected mass is larger than the total mass of the wind between the WD
and the red giant, leading to little deceleration of the shock as it
plows through the wind in the few months following the eruption.  A
nova eruption from Mira would thus perhaps look more like those of the
symbiotic slow novae than, say, an X-ray bright eruption of the
symbiotic recurrent nova RS~Ophiuchi \citep{sokoloski06} or the
$\gamma$-ray producing explosion of the symbiotic V407~Cygni
\citep{abdo10}.  From the work of \citet{townsley04}, we would expect
a nova recurrence time of $\approx 10(2)\times 10^5 \ {\rm yr}$ for
$M=0.6(1.2)M_\odot$ and $\langle \dot M \rangle =2\times
10^{-10}M_\odot \ {\rm yr^{-1}}$.

Our conclusion that Mira~B is a WD accreting at $ \dot M \sim
10^{-10}M_\odot \ {\rm yr^{-1}}$ has implications for the mode and
efficiency of mass transfer in wide binaries.  Although
\cite{mohamed07} and \cite{devalborro09} have suggested that a focused
wind could enable up to tens of percent of the donor wind to be
captured, the accretion rate needed to produce 
Mira~B's luminosity and effective temperature is just 0.1\% of the RG
wind mass-loss rate \citep{ryde00}.  It is also below the Bondi-Hoyle
accretion rate of $\mdot_{BH} \approx 10^{-8}\, \msyr
(M/0.6\,\msun)^2(7\,{\rm km\, s}^{-1}/v_\infty)^3,$ where $v_\infty$
is the relative velocity of the wind at Mira~B.  Thus we find no
evidence that either gravitational focusing or wind Roche-lobe
overflow is enhancing the accretion efficiency onto the WD in this
binary, despite the image showing a bridge of X-ray emitting material
between the two stars \citep{karovska05}.

Identifying Mira~B as a WD also has ramifications for the launch site
of the bipolar ``streams'' that have been imaged in the near- and
far-UV with $Galex$ \citep{martin07} and in the optical from the
ground \citep{meaburn09}.  Based on the behavior of other WDs that
accrete from wind-fed disks in symbiotic binaries, we speculate that
the the bipolar streams emanate from the accreting WD rather than, or
in addition to, the pulsating AGB star.  \citet{mikolajewska03} has
suggested that symbiotic stars in which the wind of the red giant is
focused toward orbital plane --- as has been seen to be the case for
Mira \citep{karovska05} --- preferentially contain large, unstable
accretion disks, which tend to produce sporadic bipolar outflows
\citep{sokoloski03}.  $Galex$ observations indicate that the streams
from Mira contain about $10^{-7}\, \msun$ of material (Don Neill,
private communication), and \citet{meaburn09} estimated that the
streams were ejected over the course of the past $\sim 1,000$~yr.  The
average rate of flow into the streams was thus roughly $\langle \dot
M_{\rm streams}\rangle \sim 10^{-10}\, \msyr$.  Given our estimate of
the accretion rate, 
Mira~B could fairly easily pump $\sim 10^{-11}\, \msyr$ into the
streams, which is roughly consistent with the observational estimate
of the mass-loss from the WD by \citet{wood02} during the $IUE$ era.
Given the large uncertainties on both the rate of accretion onto
Mira~B, the average rate of mass loss from Mira~B, and the rate of
flow in the bipolar streams, we conclude that the WD remains a viable
the launch site, and that the streams do not {\em necessarily}
originate from the RG.  Mira~AB therefore does not necessarily require
a mechanism for generating fast, bipolar outflows from the AGB star
itself, and it does not provide evidence that such mechanisms are
needed to generate bipolar symmetry in PNe.


\acknowledgments

JLS is grateful to many colleagues for useful discussions, including
J. Patterson, M. Karovska, J. Applegate, G. Luna G. Hussain,
S. Kenyon, K. Mukai, and E. Sion.  Support for this work was provided
by the National Aeronautics and Space Administration through Chandra
Award SAO G09-0027X issued by the Chandra X-ray Observatory Center,
which is operated by the Smithsonian Astrophysical Observatory for and
on behalf of the National Aeronautics Space Administration under
contract NAS8-03060, and by the National Science Foundation under
grants PHY 05-51164 and AST 07-07633.  We acknowledge with thanks the
variable star observations from the AAVSO International Database
contributed by observers worldwide and used in this research.



{\it Facilities:} \facility{Nickel, AAVSO}.




\clearpage

\clearpage





\begin{thebibliography}{}

\bibitem[Abdo et al.(2010)]{abdo10} Abdo, A. A., et al. 2010,
  Science, in press
\bibitem[Barros(2008)]{barros08} Barros, S. B. 2008, Ph. D. thesis,
  University of Warwick, Physics
\bibitem[Blondin et al.(2009)]{blondin09} Blondin, S., et al. 2009,
  \apj, 693, 207
\bibitem[Bruch(1992)]{bruch92} Bruch, A. 1992, \aa, 266, 237
\bibitem[Bruch(2000)]{bruch00} Bruch, A. 2000, \aa, 359, 998
\bibitem[Bujarrabal et al.(2001)]{bujarrabal01} Bujarrabal, V.,
  Castro-Carrizo, A., Alcolea, J., \& S{\'a}nchez  Contreras, C. 2001,
  \aa, 377, 868
\bibitem[De Marco(2009)]{demarco09} De Marco, O. 2009, \pasp, 121, 316
\bibitem[de Val-Borro, Karovska, \& Sasselov(2009)]{devalborro09} de
  Val-Borro, M., Karovska, M., \& Sasselov, D. 2009, \apj, 700, 1148
\bibitem[Done, Gierli{\'n}ski, \& Kubota(2007)]{done07} Done, C.,
  Gierli{\'n}ski, M., \& Kubota, Aya 2007, \araa, 15, 1
\bibitem[Eze, Luna, \& Smith(2010)]{eze10} Eze, R. N. C., Luna,
  G. J. M., Smith, R. K. 2010, \apj, 709, 816 
\bibitem[Froning, Long, \& Baptista(2003)]{froning03} Froning, C. S.,
  Long, K. S., \&  Baptista, R. 2003, \aj, 126, 964 
\bibitem[Gierli{\'n}ski, Niko{\l}ajuk, \&
  Czerny(2008)]{gierlinski08} Gierli{\'n}ski, M., Niko{\l}ajuk, M., \&
  Czerny, B. 2008, \mnras, 383, 741
\bibitem[Gilliland(1992)]{gil92} Gilliland, R. L. 1992, in Astronomical
  CCD Observing and Reduction Techniques, ASP Conference Series,
  Volume 23, Howell, S. B. ed., 68
\bibitem[Hoffleit(1997)]{hoffleit97} Hoffleit, D. 1997, JAAVSO, 25, 115 
\bibitem[Huggins, Mauron, \& Wirth(2009)]{huggins09} Huggins, P. J.,
  Mauron, N., \& Wirth, E. A. 2009, \mnras, 396, 1805
\bibitem[Ireland et al.(2007)]{ireland07} Ireland, M. J., et al. 2007,
  \apj, 662, 651
\bibitem[Jura \& Helfand(1984)]{jura84} Jura, M., \& Helfand,
  D. J. 1984, \apj, 287, 785
\bibitem[Karovska et al.(1997)]{karovska97} Karovska, M., Hack, W.,
  Raymond, J. \& Guinan, E. 1997, \apj, 482, L175
\bibitem[Karovska et al.(2005)]{karovska05} Karovska, M., Schlegel, E.,
  Hack, W., Raymond, J. C., \& Wood, B. E. 2005, \apj, 623, L137
\bibitem[Kastner \& Soker(2004)]{kastner04} Kastner, J. H., \& Soker,
  N. 2004, \apj, 616, 1188
\bibitem[Kennea et al.(2009)]{kennea09} Kennea, J. A., Mukai, K.,
  Sokoloski, J. L., Luna, G. J. M., Tueller, J., Markwrdt, C. B., \&
  Burrows, D. N. 2009, \apj, 701, 1992
\bibitem[Knapp \& Morris(1985)]{knapp85} Knapp, G. R., \& Morris,
  M. 1985, \apj, 292, 640
\bibitem[Knapp et al.(2003)]{knapp03} Knapp, G. R., Pourbaix, D.,
  Platais, I., \& Jorissen, A. 2003, \aa, 403, 993
\bibitem[Luna \& Sokoloski(2007)]{luna07} Luna, G. J. M., \&
  Sokoloski, J. L. 2007, \apj, 671, 741
\bibitem[Lynden-Bell \& Pringle(1974)]{lyndenbell74} Lynden-Bell, D.,
  \& Pringle, J. E. 1974, \mnras, 168, 603
\bibitem[Martin et al.(2007)]{martin07} Martin, D. C., et al. 2007,
  \nat, 448, 780
\bibitem[Matthews \& Karovska(2006)]{matthews06} Matthews, L. D., \&
  Karovska, M. 2006, \apj, 637, L49
\bibitem[Massey(1997)]{mas97} Massey, P. 1997, ``A User's Guide to CCD
  Reductions with IRAF''
\bibitem[Mauche(1991)]{mauche91} Mauche, C. W. 1991, \apj, 373, 624
\bibitem[Mauche(2002)]{mauche02} Mauche, C. W. 2002, \apj, 580, 423
\bibitem[Meaburn et al.(2009)]{meaburn09} Meaburn, J., L{\'o}pez,
  J. A., Boumis, P., Lloyd, M., \& Redman, M. P. 2009, \aa, 500, 827
\bibitem[Meier \& Kafatos(1995)]{meier95} Meier, S. R., \& Kafatos,
  M. 1995, \apj, 451, 359
\bibitem[Miko{\l}ajewska(2003)]{mikolajewska03} Miko{\l}ajewska,
  J. 2003, in Symbiotic Stars Probing Stellar Evolution, ASP
  Conf. Ser., 303, 9
\bibitem[Miszalski et al.(2009)]{miszalski09} Miszalski, B., Acker,
  A., Moffat, A. F. J., Parker, Q. a., \& Udalski, A. 2009, \aa, 496,
  813 
\bibitem[Miyamoto et al.(1991)]{miyamoto91} Miyamoto, S., Kimura, K.,
  Kitamoto, S., Dotani, T., \& Ebisawa, K. 1991, \apj, 383, 784
\bibitem[Mohamed \& Podsiadlowski(2007)]{mohamed07} Mohamed, S., \&
  Podsiadlowski, Ph. 2007, in 15th European Workshop on White Dwarfs,
  ASP Conference Series, Vol. 372, Eds. Napiwotzki, R., \& Burleigh,
  M. R., 397
\bibitem[M{\"u}rset et al.(1997)]{murset97} M{\"u}rset, U., Wolff, B.,
  \& Jordan, S. 1997, \aa, 319, 201
\bibitem[Niko{\l}ajuk, Papadakis, \& Czerny(2004)]{nikolajuk04}
  Niko{\l}ajuk, M., Papadakis, I. E., \& Czerny, B. 2004, \mnras, 350,
  L26
\bibitem[Olson \& Etzel(1995)]{olson95} Olson, E. C., \& Etzel,
  P. B. 1995, \aj, 109, 1308
\bibitem[Patat et al.(2007)]{patat07} Patat, F., et al. 2007, Science,
  317, 924
\bibitem[Patterson \& Raymond(1985)]{patterson85} Patterson, J., \&
  Raymond, J. C. 1985, \apj, 292, 550
\bibitem[Piro \& Bildsten(2004)]{piro04} Piro, A. L., \& Bildsten,
  L. 2004, \apj, 610, 977
\bibitem[Popham \& Narayan(1995)]{popham95} Popham, R., \& Narayan,
  R. 1995, \apj, 442, 337
\bibitem[Prieur et al.(2002)]{prieur02} Prieur, J. L., Aristidi, E.,
  Lopez, B., Scardia, M., Mignard, F., \& Carbillet, M. 2002, \apjs,
  139, 249
\bibitem[Psaltis, Belloni, \& van der Klis(1999)]{psaltis99} Psaltis,
  D., Belloni, T., \& van der Klils, M. 1999, \apj, 520, 262
\bibitem[Reimers \& Cassatella(1985)]{reimers85} Reimers, D., \&
  Cassatella, A. 1985, \apj, 297, 275
\bibitem[Ryde et al.(2000)]{ryde00} Ryde, N., Gustafsson, B.,
  Eriksson, K., \& Hinkle, K. H. 2000, \apj, 545, 945
\bibitem[Sahai et al.(2007)]{sahai07} Sahai, R., Morris, M.,
  S{\'a}nchez Contreras, C., \& Claussen, M. 2007, \apj, 134, 2200
\bibitem[Simon et al.(2009)]{simon09} Simon, J. D., et al. 2009, \apj,
  702, 1157
\bibitem[Shakhovskaya(1989)]{shakhovskaya89} Shakhovskaya, N. I. 1989,
  ``Solar Physics'', IAU Colloq., 121, 375
\bibitem[Shakura \& Sunyaev(1973)]{shakura73} Shakura, N. I., \&
  Sunyaev, R. A. 1973, \aa, 24, 337
\bibitem[Sokoloski(2003)]{sokoloski03} Sokoloski, J. L. 2003, JAAVSO,
  31, 89
\bibitem[Sokoloski, Bildsten, \& Ho(2001)]{sokoloski01} Sokoloski, J. L,
  Bildsten, L, \& Ho, W. C. G. 2001, \mnras, 326, 553 (SBH)
\bibitem[Sokoloski \& Kenyon(2003)]{sokokenyon03} Sokoloski, J. L., \&
  Kenyon, S. J. 2003, \apj, 584, 1027
\bibitem[Sokoloski et al.(2006)]{sokoloski06} Sokoloski, J. L., Luna,
  G. J. M., Mukai, K., Kenyon, S. J. 2006, Nature, 442, 276
\bibitem[Sunyaev \& Revnivtsev(2000)]{sunyaev00} Sunyaev, R., \&
  Revnivtsev, M. 2000, \aa, 358, 617
\bibitem[Townsley \& Bildsten(2003)]{townsley03} Townsley, D. M., \&
  Bildsten, L. 2003, \apj, 596, L227
\bibitem[Townsley \& Bildsten(2004)]{townsley04} Townsley, D. M., \&
  Bildsten, L. 2004, \apj, 600, 390 
\bibitem[Townsley \& G{\"a}nsicke(2009)]{townsley09} Townsley, D. M.,
  \& G{\"a}nsicke, B. T. 2009, \apj, 693, 1007
\bibitem[Walker(1957)]{walker57} Walker, M. F. 1957, IAU Symp., 3, 46
\bibitem[Warner(1972)]{warner72} Warner, B. 1972, \mnras, 159, 95
\bibitem[Warner(1995)]{warner95} Warner, B. 1995, ``Catclysmic Variable
  Stars'', Camb. Astrophys. Ser., Vol. 28 (Cambridge University Press)
\bibitem[Warner(2004)]{warner04} Warner, B. 2004, \pasp, 116, 115
\bibitem[Webbink, Rappaport, \& Savonije(1983)]{webbink83} Webbink,
  R. F., Rappaport, S., Savonijie, G. J. 1983, \apj, 270, 678
\bibitem[Wheatley, Mauche, \& Mattei(2003)]{wheatley03} Wheatley,
  P. J., Mauche, C. W., \& Mattei, J. A. 2003, \mnras, 345, 49
\bibitem[Wood, Karovska, \& Raymond(2002)]{wood02} Wood, B. E.,
  Karovska, M., \& Raymond, J. C. 2002, \apj, 575, 1057
\bibitem[Wood \& Karovska(2004)]{wood04} Wood, B. E., \& Karovska,
  M. 2004, \apj, 601, 502

\end{thebibliography}
\end{document}